\documentstyle[12pt]{article}
\begin{document}
\def\strut{\rule[-.5cm]{0cm}{1cm}}
\def\dspace{\baselineskip = .30in}

\title{
\begin{flushright}
{\large\bf IFUP-TH 21/95}
\end{flushright}
\vspace{1.5cm}
\Large\bf REMOVING THE COSMOLOGICAL BOUND ON THE AXION SCALE}

\author{{\bf Gia  Dvali}\thanks{Permanent address: Institute of Physics,
Georgian Academy of Sciences,  \hspace{1cm}380077 Tbilisi, Georgia.
E-mail:
dvali@ibmth.difi.unipi.it }\\ Dipartimento di Fisica, Universita di Pisa
and INFN,\\ Sezione di Pisa I-56100 Pisa, Italy\\}

\date{ }
\maketitle

\begin{abstract}The current cosmological bound on the invisible
axion scale may be avoided in the class of theories in which the gauge
coupling constant is determined through the expectation value of some
scalar field (e.g. moduli in supergravity and string theories).
This leads to the
cosmological scenario different from that of the standard invisible
axion, since the initial values of the fields are usually far away from
their true minima, allowing for the color group becoming strong
in the very early universe and fixing the axion field to its minimum.
The effect disappears as soon as
scalar field adjusts to its present value, but the above is enough to
ensure that the deviation of the axion expectation value from the minimum is
negligible at the moment of the QCD phase transition and thus to
eliminate the troublesome coherent oscillations. This may imply that
the standard axion window does not necessarily hold in generic
supergravity theories. The above observation may open a natural possibility
for the existence of the axion resulting
from the GUT or R-symmetry breaking.

\end{abstract}
\newpage

\dspace

\section{ Introduction}

  The Peccei-Quinn (PQ) mechanism [1] is a commonly accepted solution to the
strong CP problem. This mechanism is based on the concept of the
spontaneously broken anomalous chiral $U(1)_{PQ}$ symmetry, with
a subsequent light pseudogoldstone boson - axion. Phenomenological
and astrophysical constrains [2] suggest to attribute the breaking of
$U(1)_{PQ}$ to some $SU(2)\otimes U(1)$-singlet Higgs with a
large ($f_a> 10^9 GeV$ or so) vacuum expectation value (VEV) and therefore,
to make axion
invisible[3].
Furthermore, as is well known[4], the cosmological considerations
give the upper bound as well
\begin{equation}
f_a < 10 ^{12} GeV
\end{equation}
and one is left with a narrow window.

 Clearly, it is extremely important to know if one can avoid
(and at what price) the upper bound on the axion scale.

 First of all, this would give a natural possibility to implement
Peccei-Quinn mechanism in GUTs (without introducing unmotivated
intermediate scales).

Secondly, this would solve the cosmological problem of
R-axion in generic supersymmetric and supergravity theories[5], even if
R-symmetry is exact apart of anomaly. More excitingly, this could give
a possibility to use R-symmetry as $U(1)_{PQ}$.

Most importantly perhaps, we want to understand  how model
independent the cosmological bound is. In the present paper we will
argue that this bound is model dependent and in a large
class of theories may be absent. In particular this are the theories in
which the strong gauge coupling constant is determined by an expectation
value of a scalar field, e.g. generic supergravity and
superstring theories.

\dspace

\section{ Cosmological Constraint}

  Let us first briefly recall the origin of the constraint. In the
invisible axion scenario[3], the breaking of the Peccei-Quinn
symmetry is induced by the $SU(3)\otimes SU(2)\otimes U(1)$-singlet
Higgs field $\phi$ with a `Mexican hat' potential

\begin{equation}
V=\lambda(|\phi|^2 - f_a^2)^2
\end{equation}

At the bottom of this potential the complex field can be represented
as $\phi =f_ae^{i{a \over f_a}}$, where $a$ is the angular
goldstone mode - axion, parameterizing the flat minimum.
QCD instanton effects lift the vacuum degeneracy
and induce effective potential for $a$
\begin{equation}
V_a = \Lambda_{QCD}^4(1 - cos{aN \over f_a})
\end{equation}
Here $N$ stands for the non-anomalous $Z_N$ subgroup and may lead to the
domain wall problem [6] if the Peccei-Quinn phase transition (if it happens
$at~all$ [7]) takes place after inflation. In our scenario this
never happens (essential point is that PQ field is nonzero and large
during and after inflation) and all topological defects are inflated away. So
below we will simply assume $N = 1$.

In the context of the standard big bang
scenario it is usually assumed that the phase transition with
$U(1)_{PQ}$-symmetry breaking occurs when the universe cools below
the temperature $T_c \sim f_a$. Certainly, this is so for the case of
a single Higgs field, but need not be true in general. First, as it
was shown recently [7], even the minimal invisible axion model can
exhibit symmetry $nonrestoration$[8] at high $T$. Secondly,
in the inflationary scenarios the VEV
of $\phi$ can be strongly shifted during inflation due to the coupling with
the inflaton field and therefore be nonzero from the `very beginning'. (In
general, this is true for any Higgs field, so
that even $SU(2)\otimes U(1)$ can
be strongly broken during inflation).  One way or another, in the standard
case the crucial assumption is that
from the very moment of the Peccei-Quinn phase transition and all the way
down to the temperatures $\sim \Lambda_{QCD}$, the bottom
of the potential (2) is exactly flat and there is no prefered value
of $a$ during this period ((3) vanishes). Consequently, at the moment of the
QCD phase transition, when the instanton effects lift degeneracy,
$a$ rolls to the minimum and starts coherent oscillations about it
with large initial amplitude $A \sim f_a$. The energy density stored
in such oscillation is
\begin{equation}
     E_a \sim  m_a^2 A^2
\end{equation}
where $m_a \sim {\Lambda_{QCD}^2 \over f_a}$
is the axion mass. Of course,
the switch-on of the axion mass is not sudden
and this fact somewhat reduces the constraint.
More detailed analysis [4] shows that universe had to be overclosed by
above coherent oscillations unless the VEV of $\phi$ is restricted
by (1).

Clearly, this upper bound on $f_a$ can be avoided, if there
have been some mechanism guaranteeing that at the moment of QCD phase
transition $a$ starts oscillations with much smaller initial value
$A << f_a$. This can be the case if in the early universe one assumes
a period during which vacuum degeneracy was strongly lifted, with
axion having a mass of order the Hubble constant ($H$), so that
$a$ could rapidly settle in to the minimum $a = 0$.
Of course, below certain temperature $T_R >> \Lambda_{QCD}$ this
effect had to disappear and bottom of the potential had to become
again flat until the `usual' QCD phase transition, but the dramatic
consequence of such a period would be that at $T_R$ axion appears to
sit at its zero temperature minimum.  Below $T_R$ the thermal
fluctuations will try to drive $a$ away from zero, but resulting
deviation for the moment of QCD transition will be very small
if $T_R$ is small enough. In the next section we consider
theories exhibiting such a cosmological behavior.

\section{ Scenario with inflation}

The theory with above nonstandard cosmological history can be one
in which the gauge coupling constant
is fixed by VEV(s) of some scalar field(s). For example,
such a situation is common in generic supergravity and superstring
theories in which the gauge coupling constants is determined through
the following type term in the Lagrangian [9]

\begin{equation}
  f({Z \over M_p})F_{\mu \nu}F^{\mu \nu}
\end{equation}

where $F_{\mu \nu}$ is gauge field tensor, $f$ is some (real) function
of the field $Z$ and $M_p$ is the Planck scale.
In the superstring theories $Z$ is a dilaton field ($S$) and
function $f$
has the form (in Planck scale units) $f(S) = ReS + thresold~corrections$.
In generic supergravity theories $f(Z)$ may have whatever form, provided
that its low temperature
expectation value sets the correct gauge
coupling constant in the true vacuum with zero energy.
Certainly, there is no reason to assume that the value $f$ had to be the
same in the early universe, since the scalar VEVs, in general, are
far away (by $\Delta Z \sim M_p$) from their low temperature minimum.
This is the case both because of
thermal and quantum fluctuations and also because the minimum itself becomes
displaced from the present place. In particular, this fact is precisely
the origin of the cosmological moduli problem [10], and it is interesting
how the assumption that creates a problem in one case, may provide
a solution in another. For us the most important consequence of this
fact is that it can allow for the QCD becoming strong during some
period in the early history
and giving to the axion the mass $\sim$ Hubble constant at that time.

Most easily (but not only) this may happen in the inflationary
scenarios[11], since both
$Z$ and $\phi$ in general are getting large displacement due to interaction
with the inflaton field. For example, if $Z$ is a string moduli, in general
it will get large $\sim M_P$ shift due to the fact that normally inflation
generates a curvature $\sim H$ to its potential [12], and because at present
potential is very flat (and VEV is $\sim M_p$), the resulting
shift is very large. In addition
it is also possible that $Z$-field, that fixes gauge coupling, itself
can be an inflaton field.

In general VEV of the PQ field also will be displaced from its true
value $f_a$. If selfcoupling of $\phi$ is not very small, the mass of the
radial mode at the bottom of the potential is $\sim f_a$. Since in the
present context we assume $f_a$ being non-standardly large (GUT scale or
even $M_p$),
significant displacement is not expected, unless $\phi$ is directly coupled
to the inflaton (or unless $H > f_a$, which is unlikely to be the case).
Of course, it is perfectly possible that either selfcoupling is very small,
or PQ field has unsuppressed coupling with inflaton. In such a case
the displacement can be large. So we assume that the natural nonzero VEV
of $\phi$ in inflationary epoch is $\sim f_a$ (or even larger).

Now what about the early history of the electroweak Higgs sector? Again,
this is very model-dependent situation. Behavior of the electroweak
Higgs doublets in the inflationary era is determined by several
factors and in particular by their coupling with other (GUT) Higgses
and with inflaton. However, even in the minimal cases their displacement
from the normal (almost zero) VEVs can be large. For example,
in the minimal supersymmetric standard, model with a single pair of
the Higgs doublets $H$ and $\bar H$, the scalar potential is almost
flat in the direction $|H| = |\bar H| = arbitrary$
and $all~other~fields = 0$.
(There are other flat directions which involve squarks and slepton
VEVs, but they are not of our interest here).
In the supersymmetric limit this flat direction is
shut down by the supersymmetric mass term $\mu H\bar H$ (
$\mu$-term) in the superpotential, which generates small weak scale
curvature ($\sim 100 GeV$). After SUSY breaking there is an additional
contribution of the comparable strength from the soft terms.
Therefore, effective curvature in $SU(2)\otimes U(1) - breaking$ direction
is very small and resulting shift during inflation can be large.
In the supergravity case, even in the absence of the direct
coupling with inflaton, this direction will get universal
gravity transfer contribution $\sim H$ to the curvature,
very much like moduli fields.
For the canonical Kahler potential this contribution is positive,
but can be negative for the more general form. However, unlike
moduli, Higgs doublets are gauge nonsinglets and have renormalizable
couplings with matter fields and they
can (and will) get negative corrections to the mass from other (radiative)
sources as well [13]. So the outcome is that, in general, there is
no particular reason for the Higgs doublet VEVs to vanish during
inflation and their typical magnitude can be at least $\sim H$.

One way or another, in general we expect that
QCD may naturally become strong for some period  (during inflation).
Assuming that at that time $m_a \sim H$, $a$ will settle at $a = 0$ and
stay there before QCD instantons will be switched-off.
 Clearly, this effect has to disappear after inflation ends (or may even
before) and all the VEVs ($\phi,Z,H,\bar H$) adjust to their normal values.
Below this point universe is reheated to some temperature $T_R$ and
the hot bing bang starts. Influence of thermal effects on the values
$\phi$ and $f({z \over M_p})$ are negligible provided $T_R << f_a,M_p$
and so we assume that in the interval of temperatures
$T_R > T > \Lambda_{QCD}$ the QCD instantons are switched-off
and $m_a = 0$, as it is
assumed in the standard scenarios. The important outcome in our case
is that at $T \sim T_R$ axion field is fixed at $a =0$. But now, the thermal
fluctuations will try to drive $a$ away from zero. Since the
bottom of the potential is exactly flat, journey along it can be considered
as a random walk with a step $\Delta a \sim T$ per Hubble time. This means
that during the temperature interval $T_R - \Lambda_{QCD}$ it can cause
(at best) the deviation $A \sim T_R$ of the axion field from the minimum.
Thus the energy stored
in the coherent oscillations becomes
\begin{equation}
 E_a \sim m_a^2 T_R^2 \sim \Lambda_{QCD}^4 {T_R^2 \over f_a^2}
\end{equation}
We see that this energy is suppressed with respect to the standard case
by an extra factor $\sim {T_R^2 \over f_a^2}$.
No definite information exists about the reheat
temperature $T_R$ at present. For
example, the solution of the gravitino problem requires $T_R < 10^9$[14].
In general, if the electroweak baryogenesis [15] can be trusted, $T_R$
can be as small as electroweak scale. This means that for the GUT
scale axion the oscillation energy may be suppressed by a factor
of order $10^{-28}$ with respect to the usual case! For our purposes
such a small reheat temperature is not needed and it is perfectly
enough that $T_R$ is few orders of magnitude below $f_a$.
This fact avoids
cosmological bound (1) on the axion scale and allows it to be
as large as the GUT or Planck scale without any trouble.

\section{Quantum fluctuations during inflation}

In the above analysis, for simplicity, we were assuming that the epoch
 of the
superstrong QCD ends together with inflation. Of course, in a more
general case this may happen while inflation is still in progress.
Since axion potential becomes flat, the quantum de Sitter fluctuations,
presented during inflation,
will drive $a$ away from zero very much like thermal effects in
the previous section. Now the step is $\Delta a \sim H$ during the
Hubble time $H^{-1}$ and resulting dispersion after $N$ Hubble times
is $ \sim H \sqrt N$[16]. Anisotropy of the microwave background
radiation induced by the fluctuations in the axion field at the late
stages of inflation may require small (large) values of $H$ ($f_a$)
at that time [17].
After reheating, the contribution of the thermal fluctuations
to the amplitude is similar as in our previous discussions and
is of order $T_R$.
Of course, the precise values $H$ and $T_R$ are model-dependent and
can not be studied here, but the message is that in any case
coherent oscillations can be strongly suppressed, if both
$T_R$ and $H$ (during inflation) are somewhat smaller $f_a$.

\section{Constraints from axionic cosmic strings and domain walls}

Another possible constraint[18] on the Peccei-Quinn scale comes
from the decay of the global axionic strings [19] and the
models with $N > 1$ suffer from the domain wall problem[6].
It should be clear
that no such constraint can be applied in our case, since
essential point is that both $f_a$ and $m_a$ are nonzero and large
during inflation. This simply means that all
existing topological structures such as string-wall systems[19] and
walls without strings (pure strings can not exist due to nonzero
$m_a$) will be inflated away. Production of the new defects
by quantum de Sitter fluctuations is exponentially suppressed
by the factors $\sim {f_a^2 \over H^2}$ and $\sim {m_af_a^2 \over H^3}$
for the string and wall cases respectively [20].

\section{Anthropic principle}

Number of authors [21] have pointed out that the cosmological
constraint (1) can be avoided by the arguments based on the anthropic
principle. According to this arguments we observe universe with
$A << f_a$, because inflation has produced domains filled with all
possible values of the axion field $a$ ($A$), but in the domains
with $A \sim f_a$ life is impossible. Therefore, the only place we
have a chance to see ourselves is a domain with small enough $A$.

In our scenario there is no need in anthropic principle, since now,
all the domains produced by inflation have $A << f_a$ due to the
fact that $m_a$ was large in that epoch.

\section{Summary and outlook}

We have argued that there is a large class of theories in which
current cosmological constraints on the axion scale may
be naturally avoided. In particular, this are theories in
which a VEV of a scalar field can determine the strength of
the gauge coupling, as it happens in generic supergravity and
superstring theories. In such an approach the gauge constant becomes
a dynamical variable and is driven by the evolution of the scalar
field. The crucial point is that in the early
universe VEVs are far away from their true minima, allowing
for QCD  to become strong at some high scale and fixing axion
field to minimum. The deviation from $a = 0$ (induced by thermal
or quantum fluctuations) at the moment of the
ordinary QCD phase transition is determined by the reheating
temperature and can be naturally small.

 This effect opens a possibility for the existence of a
GUT (or $M_p$) scale axion and for avoiding problems related
with the R-axion in supersymmetric theories.

\section*{Acknowledgments}

I would like to thank G.Senjanovic for very useful discussions and
suggestions about the possible alternative schemes. I would like to
gratefully acknowledge useful conversations with R.Barbieri
Z.Berezhiani and M.Shaposhnikov.

  \section*{References}

\begin{enumerate}

\item R.D.Peccei and H.R.Quinn, {\it Phys.Rev}, D16 (1977) 1791.

\item For a review see J.E.Kim, {\it Phys.Rep.}, 150 (1987) 1.

\item  J.Kim, {\it Phys.Rev.Lett}, 43 (1979) 103;
M.A.Shifman, A.I.Vainshtein and V.I.Zakharov, {\it Nucl.Phys.}
B166 (1980) 493; M.Dine, W.Fischler and M.Srednicki, {\it Phys.Lett.}
B104 (1981) 199; A.P.Zhitnitskii, {\it Sov. J. Nucl.} 31
(1980) 260.

\item J.Preskill, M.Wise and F.Wilczek, {\it Phys.Lett.} B120 (1983)
127;
      L.Abbot and P.Sikivie, {\it Phys.Lett.} B120 (1983) 133;
    M.Dine and W.Fischler, {\it Phys. Lett.} B120 (1983) 137.

\item  T.Banks, D.Kaplan and A.Nelson, {\it Phys.Rev.}, D49 (1994) 779.

\item P.Sikivie, {\it Phys.Rev.Lett} 48 (1982) 1156.

\item  G.Dvali and G.Senjanovic, Preprint IFUP-TH 61/94 (to be published).

\item S.Weinberg, {\it Phys.Rev.}, D9, (1974) 3357.
R.N.Mohapatra and G.Senjanovi{\'c}, {\it Phys.Rev.Lett},
42, (1979) 1651; {\it Phys.Rev.} D20, (1979) 3390.

\item E.Cremmer, S.Ferrara, L.Girardello and A.Van Proeyen, {\it Nucl.Phys},
B212 (1983) 413;

H.P.Nilles, {\it Phys.Rep}, 110 (1984) 1.

\item G.Coughlan, W.Fischler, E.Kolb, S.Raby and G.Ross, {\it Phys.Lett},
B131 (1983) 59;
J.Ellis, D.V.Nanopoulos and M.Quiros, {\it Phys.Lett.}, B174 (1986) 176;
G.German and G.G.Ross, {\it Phys.Lett.}, B172 (1986) 305;
O.Bertolami, {\it Phys.Lett}, B209 (1988) 277; R. de Carlos, J.A.Casas,
F.Quevedo and E.Roulet, {\it Phys.Lett.}, B318 (1993) 447.

\item A.D.Linde, {\it Particle Physics and Inflationary Cosmology}
(Harwood Academic, Switzerland, 1990); E.W.Kolb and M.S.Turner,
{\it The Early Universe} (Addison-Wesley, Reading, MA, 1990).

\item G.Dvali, Preprint IFUP-TH 09/95, hep-ph/9503259; IFUP-TH 10/95,
hep-ph/9503375 (to be published in Phys.Lett.B);

     M.Dine, L.Randall and S.Thomas, Preprint hep-ph/9503303.

See also,
M.Dine, W.Fischler, and D.Nemeschansky, {\it Phys.Lett} B136
(1984) 169 and in another context by O.Bertolami and G.Ross,
{\it Phys.Lett.} B183 (1987) 163.
E.Copeland, A.R.Liddle, D.H.Lyth, E.D.Stewart and D.Wands,
{\it Phys.Rev}, D49 (1994) 417.

\item Since moduli have no renormalizable interactions, gravity is the
only messenger that can generate their masses $\sim H$ during inflation.
However, for the gauge nonsinglet flat directions (squarks, sleptons,
Higgs doublets) inflation induced SUSY breaking can be equally  well
transmitted radiatively, by gauge and/or Yukawa interaction. In fact,
gauge loops induce universal (up to a gauge quantum number) curvature
to all flat directions, if the inflaton $F$-term is coupled to (at least)
some gauge nonsinglets.  Yukawa couplings can also drive negative
$[mass]^2$ of $H,\bar H$. For a more detailed discussion see,
G.Dvali, Preprint IFUP-TH 10/95, hep-ph/9503375.

\item J.Ellis, A.D.Linde and D.Nanopoulos {\it Phys.Lett.}, B118,
(1982) 59;
D.Nanopoulos, K.Olive and M.Srednicki,
{\it Phys.Lett.}, B127, (1983) 30;
W.Fischler, {\it Phys.Lett}, B332 (1994) 277.

\item For the earlier work on this subject see, e.g.,
V.A.Kuzmin, V.A.Rubakov and M.E.Shaposhnikov, {\it Phys.Lett.}
B155 (1985) 36; Shaposhnikov, {\it Nucl.Phys.} B287 (1987) 757; B299
(1988) 797.

For a review see, e.g. A.G.Cohen, D.B.Kaplan and A.E.Nelson,
{\it Ann.Rev. of Nucl. and Part.Phys.} 43 (1993) 27.

\item A.D.Linde and D.H.Lyth, {\it Phys. Lett.}, {\bf B246}, (1990) 353.

\item M.S.Turner and F.Wilczek, {\it Phys.Rev.Lett} 66 (1991) 5.

Models of inflation avoiding this constraint were discussed by
A.Linde, {\it Phys.Lett.} B259 (1991) 38.

\item R.L.Davis and E.P.S.Shellard, {\it Nucl.Phys.}, B324 (1989) 167.

\item A.Vilenkin and A.Everett, {\it Phys.Rev.Lett.} {\bf 48}
(1982) 1867.

\item R.Basu, A.H.Guth and A.Vilenkin, {\it Phys.Rev.}, D44 (1991)340.

\item  S.-Y.Pi, {\it Phys.Rev.Lett.} 52 (1984) 1725; M.S.Turner,
{\it Phys.Rev.} D33 (1986) 889; A.D.Linde, {\it Phys.Lett.} B201
(1988) 437.

\end{enumerate}
\end{document}